\documentclass[conference]{IEEEtran}
\IEEEoverridecommandlockouts
\usepackage{cite}
\usepackage{amsmath,amssymb,amsfonts}
\usepackage{algorithmic}
\usepackage{graphicx}
\usepackage{textcomp}
\usepackage{xcolor}
\usepackage{url}
\usepackage{cite}
\usepackage{pifont}
\usepackage{microtype}
\usepackage{tikz}
\usepackage{hyperref}
\usepackage{lipsum}
\usepackage[a4paper, total={184mm,239mm}]{geometry}
\def\BibTeX{{\rm B\kern-.05em{\sc i\kern-.025em b}\kern-.08em
    T\kern-.1667em\lower.7ex\hbox{E}\kern-.125emX}}

\newcommand\copyrighttext{%
  \footnotesize \textcopyright 2025 IEEE. Personal use of this material is permitted.
  Permission from IEEE must be obtained for all other uses, in any current or future
  media, including reprinting/republishing this material for advertising or promotional
  purposes, creating new collective works, for resale or redistribution to servers or
  lists, or reuse of any copyrighted component of this work in other works.
  DOI: \href{https://doi.org/10.23919/DATE64628.2025.10992734}{10.23919/DATE64628.2025.10992734}}
\newcommand\copyrightnotice{%
\begin{tikzpicture}[remember picture,overlay]
\node[anchor=south,yshift=10pt] at (current page.south) {\fbox{\parbox{\dimexpr\textwidth-\fboxsep-\fboxrule\relax}{\copyrighttext}}};
\end{tikzpicture}%
}

\begin{document}

\newcommand{\RomanNumeralCaps}[1]{\MakeUppercase{\romannumeral #1}}

\title{Improving Chip Design Enablement for\\Universities in Europe -- A Position Paper
\thanks{This work is part of the Chipdesign Germany project, funded by the German Federal Ministry of Educaton and Research (BMBF) under grant no. 16ME0890.

This work also received support from the French National Research Agency (ANR) under France 2030 with grant no. ANR-22-PEEL-0013 (PEPR \'Electronique - CHOOSE).}
}

\author{\IEEEauthorblockN{Lukas Krupp\IEEEauthorrefmark{1}, Ian O'Connor\IEEEauthorrefmark{2}, Luca Benini\IEEEauthorrefmark{3}\IEEEauthorrefmark{4}, Christoph Studer\IEEEauthorrefmark{3}, Joachim Rodrigues\IEEEauthorrefmark{5} and Norbert Wehn\IEEEauthorrefmark{1}}
\IEEEauthorblockA{\IEEEauthorrefmark{1}RPTU University of Kaiserslautern-Landau, \IEEEauthorrefmark{2}\'Ecole Centrale de Lyon}
\IEEEauthorblockA{\IEEEauthorrefmark{3}ETH Zurich, \IEEEauthorrefmark{4}University of Bologna, \IEEEauthorrefmark{5}Lund University}
}

\maketitle
\copyrightnotice

\begin{abstract}
The semiconductor industry is pivotal to Europe’s economy, especially within the industrial and automotive sectors. However, Europe faces a significant shortfall in chip design capabilities, marked by a severe skilled labor shortage and lagging contributions in the design value chain segment. This paper explores the role of European universities and academic initiatives in enhancing chip design education and research to address these deficits. We provide a comprehensive overview of current European chip design initiatives, analyze major challenges in recruitment, productivity, technology access, and design enablement, and identify strategic opportunities to strengthen chip design capabilities within academic institutions. Our analysis leads to a series of recommendations that highlight the need for coordinated efforts and strategic investments to overcome these challenges.
\end{abstract}

\begin{IEEEkeywords}
semiconductor industry, chip design, Europe, universities, design enablement
\end{IEEEkeywords}

\section{Introduction}
The semiconductor industry is central to the European economy, particularly in the industrial and automotive sectors \cite{b9}. The required types of semiconductors are very heterogeneous, ranging from analog, radio-frequency (RF), and power management chips to sensors and specialized processors \cite{b10}. Europe covers 55\% of the global market within these areas. Furthermore, the European industry significantly contributes to equipment and materials for semiconductor production, accounting for 40\% and 20\% of the respective segments in the semiconductor value chain \cite{b8}. Semiconductor fabrication and chip design are the two largest segments of the value chain, with 34\% and 30\% share of added value, but Europe only contributes 8\% and 10\% respectively to these two segments \cite{b1}. Therefore, Europe is currently strengthening its semiconductor fabrication capacity through significant investments, exemplified by new fabs from TSMC in Dresden, Germany; STMicroelectronics in Catania, Italy; and Intel in Leixlip, Ireland. Yet, there remains a substantial deficit in chip design capabilities in Europe, worsening in the future if no actions are taken \cite{b1, b2}.

The deficit in the European chip design capability manifests in a skilled labor shortage. The EU MicroElectronics Training Industry and Skills (METIS) Report 2023 has shown that chip designers are the job profiles identified as the most difficult to find and the second-most needed in the European microelectronics industry \cite{b3}. European universities are crucial stakeholders in solving this problem, as they educate the next generations of chip designers. Yet, as in other countries, such as the United States \cite{b4}, universities in Europe face political, educational, legal, and technical obstacles holding them back in the race to overcome the skilled labor shortage in chip design.

The specializations of research groups at European universities align well with the heterogeneity of the chip design industry. This alignment is reflected in the application areas and the technologies used \cite{b6,b7}. However, the chip design capabilities of European universities extend far beyond that. Opportunities arise with novel computing paradigms like neuromorphic computing, new devices like resistive RAM (RRAM), integration techniques like chiplets, or design methodologies based on artificial intelligence (AI), to name a few. Universities have the potential not only to address the skilled designer shortage but also to drive innovation, enabling expansion into new market segments and further increasing Europe’s market share in chip design. This untapped potential positions universities as key enablers for the future growth of chip design in Europe.

The focus of this position paper is on chip design within academic education and research, providing a perspective from European university and academic initiative representatives. We aim to give an overview of European academic initiatives for strengthening chip design at universities and assess the challenges and opportunities in this domain. We provide recommendations to overcome existing challenges and improve the enablement of universities in chip design. We emphasize the following topics:
\begin{itemize}
    
    \item \textbf{Recruiting and Education}: Interest among students in software engineering and AI is growing, while interest in microelectronics is declining. The number of graduates in semiconductor-related fields has stagnated at the European level and even declined in some countries over the past years \cite{b2}. Efforts to attract more students in Europe do not yet address the entire pipeline from early engagement with children and high-school students to university students in semiconductor-related fields who have not yet specialized. Furthermore, job and salary opportunities in software engineering and AI are perceived as more attractive than those in semiconductor-related fields.

    \item \textbf{Productivity}: Software engineering and AI are productivity-driven and only minimal infrastructure is necessary to become productive (fast road to success). For example, a single line of Python code can generate thousands of assembly instructions. In contrast, chip design entails a longer path to success. Unlike software engineering, hardware design is primarily efficiency-driven. Before being productive, a range of specialized tools and design knowledge have to be acquired. Setting up a design flow for a specific technology node is challenging and the corresponding electronic design automation (EDA) tools are difficult to learn and use. Hence, in teaching, maintaining student motivation is demanding, as the path to visible success is long. In research, the time to set up the necessary infrastructure and acquire the knowledge to perform advanced design detracts from exploration and innovation. Furthermore, the relatively short duration of doctoral studies results in frequent changes of staff and necessitates continual re-training of new students. 

    \item \textbf{Technology, Cost, and Law}:  Complicated legal agreements such as non-disclosure agreements (NDAs) as well as export control restrictions based on students' countries of origin or visa statuses, limit the access to process design kits (PDKs), libraries, intellectual property (IP) blocks, and state-of-the-art EDA tools. Further, these restrictions hinder collaborative design efforts. Moreover, the costs for multi-project wafer (MPW) runs for academia in Europe are still high, especially for advanced technology nodes, and are often not covered by research funding. The turn-around times from design to packaged chips also exceed typical course lengths, thesis or research project durations.
    
    \item \textbf{Availability vs.~Enablement}: 
    Europractice, the most prominent platform for MPW services among universities in Europe \cite{b5}, provides access to commercial state-of-the-art EDA tools, advanced technologies, standard-cell libraries, and IP. We refer to the provision of this data and tools as {\em availability}. However, as previously mentioned, setting up a complete design flow and performing the configuration for a given technology requires expert knowledge and experience. We denote the tasks required to make the transition from availability to being able to carry out a design with a state-of-the-art flow as {\em enablement}.
\end{itemize}

The rest of the paper is structured as follows. Section~\RomanNumeralCaps{2} presents the current state of European academic initiatives that aim to strengthen chip design education and research at universities to provide a thorough evaluation of the current landscape. Section~\RomanNumeralCaps{3} addresses key challenges European universities face, focusing on the four problem areas introduced above. Section~\RomanNumeralCaps{4} identifies opportunities to overcome the key challenges and presents strategies for leveraging these opportunities. Section~\RomanNumeralCaps{5} summarizes and concludes the paper.

\section{Current State of Academic Chip Design Initiatives in Europe}
Various European and national structural initiatives aimed at industrial or academic sectors seek to stimulate innovation, competitiveness, and technological sovereignty. They are based on three fundamental axes: training and skill development for specialized talents, access to chip design tools, and access to foundries, MPW runs in advanced technologies, and packaging solutions for proof-of-concept designs. 

National initiatives, such as SwissChips\cite{swisschips}, Chips-IT \cite{chips-it}, Chipdesign Germany\cite{chipdesign-germany}, PEPR Electronique \cite{pepr-electronique}, or ClassIC \cite{class-ic}, primarily aim to energize a regional innovation network with international ambitions, support corporate innovation projects, and leverage the technological expertise of the ecosystem. They promote synergies between academic institutions, companies, infrastructures, and technologies of local research laboratories. Furthermore, they aim to strengthen the professional training system and develop a network of universities, research centers, and companies conducive to innovation and technology transfer. These initiatives are part of a European dynamic aimed at reinforcing the continent's independence and competitiveness in the semiconductor industry.

The launch of the EU Design Platform within the framework of the EU Chips Act \cite{eu-chips-act} aims to amplify these efforts with two new axes: the creation of a robust network as a central exchange platform and access to ``Design Enablement Teams'' (DET) that will offer a set of services (design environment, technical support, e.g., with design flow and PDKs, design expertise, manufacturing routes, etc.) to encourage the emergence of fabless companies and the design of next-generation integrated circuits (ICs). It specifically targets startups and SMEs.
Under the umbrella of the EU Chips Act, many EU member states will host a Chips Competence Center (CCC). The CCCs create an interface between the businesses and the infrastructure of the EU Design Platform.

Another promising direction promoted by several European academic groups is increasing the emphasis on open-source hardware \cite{fossi-roadmap}. Academic projects, such as the parallel ultra-low-power (PULP) platform \cite{pulp}, have demonstrated the feasibility of building a rich and widely reusable library of complex digital IPs, including processor cores based on the open RISC-V instruction set architecture. These IPs, released under a liberal license, have enabled the creation of a research and innovation ecosystem where interactions are not encumbered by complex confidentiality and NDAs, while at the same time enabling reproducible benchmarking. Hence, open-source hardware contributes to improved design enablement and eases the integration into education and research.

\section{Key Challenges for Universities in Chip Design}
\subsection{Recruiting and Education} \label{subsec:recruiting}
Chip design is a domain of extraordinary technical complexity that demands systematic skill-building, and early specialization. This requirement fundamentally conflicts with contemporary educational paradigms that encourage students to maintain broad career options. Middle and high-school curricula — already constrained by limited instructional hours — predominantly focus on broad scientific fundamentals rather than specific engineering domains. Such generalist approaches inadvertently fail to spark early curiosity among potential future chip designers. The result is a persistent talent shortage, where students remain unaware of the intellectual richness and societal importance of chip design until potentially too late in their academic journeys.

To counteract these structural educational barriers, various organizations are developing targeted strategies. Summer programs and student contests, such as those organized by IEEE SSCS (PICO - Platform for IC design Outreach), represent promising mechanisms for early engagement \cite{pico, invent-a-chip, semi-chipquest}. Initiatives, such as TinyTapeout \cite{tinytapeout}, which provide high school students with hands-on, introductory-level chip design experiences, also offer a pragmatic approach to stimulating interest.

Moreover, social factors are intertwined with the educational challenges previously described. Limited exposure to the field in early education not only hampers later engagement but also perpetuates misconceptions about what chip design entails \cite{vde-image-study}. These misunderstandings reduce the perceived attractiveness and can deter students from pursuing a path in chip design. Frequent aspects of misconception include Europe's capabilities and potential in chip design. The European semiconductor sector, while robust, remains largely underrecognized among the general population. This lack of visibility leads students to mistakenly assume that job opportunities in chip design are not available within their region. Furthermore, students often have little understanding of the professional life of a chip designer and it is not well communicated that careers in chip design are financially rewarding. Without insights into the profession, students cannot make informed educational choices. Uncertainty about the steps to become a chip designer poses another barrier. Students frequently are unaware of what they need to study or which academic programs will equip them with the requisite skills. Chipdesign Germany \cite{chipdesign-germany}, for example, addresses the issue of promoting young talent and students in chip design through a dedicated university alliance and a working group.

Another concern within the field of chip design is the gender and diversity gap. Insufficient action has been taken to attract and retain women and other underrepresented minorities in semiconductor engineering. This lack of diversity not only limits the range of new ideas within the industry but also perpetuates stereotypes that can deter the participation of a broader range of students from considering a career in chip design.

Political factors also influence the challenges faced in cultivating a robust talent pipeline for chip design. Investment priorities vary notably between regions, impacting the resources available for education and workforce development. For example, the United States' CHIPS Act (short for Creating Helpful Incentives to Produce Semiconductors) allocates \$200 million to the ``Chips for America Workforce and Education Fund.'' This investment underscores a strategic commitment to bolstering the domestic semiconductor industry by nurturing a skilled workforce.
In contrast, the European Chips Act proposes the establishment of ``Centers of Excellence'' to form an educational network across Europe. While this initiative aims to enhance collaboration and knowledge sharing, it lacks an explicit budget dedicated to workforce development \cite{b8}. The absence of targeted funding may limit the effectiveness of these centers in addressing the talent shortage. Similarly, SwissChips \cite{swisschips}, a recent chip-design initiative in Switzerland, 
is solely supporting chip design research in academia and does not foresee any involvement of the semiconductor industry. Therefore, its effectiveness is limited to research activities and can by far not compete with larger initiatives that strongly involve the semiconductor industry. 

Furthermore, some European countries have only recently started investing in the practical education of chip designers. Historically, there was an absence of semiconductor education as a priority on national political agendas. Hence, the full potential of the EU in chip design still remains untapped. However, the situation is beginning to change with increased EU funding aimed at promoting the semiconductor industry, alongside the rise of the open-source hardware movement. 

\subsection{Productivity} \label{subsec:productivity}
While software development benefits from high productivity facilitated by accessible and minimal infrastructure and tools, chip design remains a field where productivity is impeded by complexity, specialized knowledge requirements, and substantial infrastructural demands. To understand the inherent complexity of chip design, one can examine a typical digital application-specific IC (ASIC) design flow. The following considerations apply similarly to analog design. The design flow is divided into two main phases: frontend and backend design. The frontend encompasses the process from design specification to the generation of a verified netlist, while the backend covers the transformation of this netlist into a manufacturable chip, culminating in the creation of a GDSII file. Accordingly, we differentiate between frontend and backend productivity.

Frontend productivity focuses on how efficiently a chip designer can progress from system specifications and requirements to a Register-Transfer Level (RTL) design, from which a netlist can be synthesized. One of the primary problems in achieving high frontend productivity is the relatively low abstraction level of RTL code. A single line of RTL code typically generates only 5 to 20 gates. Furthermore, the semantic gap present in the synthesis process hinders frontend productivity. Multiple high-level descriptions in the logic design stage can lead to equal simulation behavior but produce different underlying physical implementations. These variations can have substantial impacts on the performance, power, and area (PPA) metrics of the final chip. Consequently, chip design requires more manual code writing and optimization than software development. 

Addressing these challenges requires tools and methodologies that enhance frontend productivity. High-Level Synthesis (HLS) tools allow designers to write code in high-level languages, which can then be automatically translated into RTL code, reducing the manual coding effort and bridging the semantic gap. Hardware construction languages like Chisel HCL provide powerful abstractions for hardware design, enabling more efficient code reuse and modularity. Additionally, soft IP core management platforms such as LiteX facilitate the integration and management of pre-designed components.

Backend productivity pertains to the efficiency with which a chip designer can transform a netlist into a manufacturable chip while considering the physical aspects necessary to meet PPA constraints. This phase involves multiple tools for the steps of the design flow. Additionally, the backend process is heavily influenced by the targeted technology nodes and the associated PDKs and libraries. Ensuring backend productivity necessitates the automation of the tool flow through scripting and its configuration to accommodate the technology-dependent artifacts. Performing these tasks requires significant domain-specific knowledge and can be a substantial barrier to productivity.

While Field-Programmable Gate Arrays (FPGAs) offer an alternative for digital design, they only partially cover the design flow. FPGAs are useful for prototyping but fall short in providing insights into the full backend design process required for ASIC development. Analog design lacks viable alternatives like FPGAs. Tasks such as component sizing or manual layout demand meticulous attention and cannot be easily automated, exacerbating both frontend and backend productivity challenges.

The productivity hurdles in chip design have significant implications for education and research at universities. Maintaining student motivation is challenging due the substantial upfront effort required to become productive. Acquiring the expertise to perform advanced chip designs and establishing the necessary infrastructure consumes considerable time and resources. These demands may reduce the attractiveness of chip design study or PhD programs. Additionally, setting up new courses or degree programs with a focus on chip design is daunting due to logistical challenges of acquiring and maintaining the necessary tools and PDKs. Universities must navigate licensing agreements and NDAs and provide adequate hardware resources.

\subsection{Technology, Cost, and Law} \label{subsec:technology}
The prohibitive costs associated with commercial tools and production-ready designs, which can range from \$\,5 million for a 130\,nm chip to \$\,725 million for a 2\,nm chip, put practical chip design experience out of reach for most educational institutions. This financial barrier is compounded by the complexity of obtaining access to foundries and test infrastructure. Export control and NDAs further complicate the situation, limiting certain students' access to advanced technologies.

To address the cost-related challenges, EDA companies have, for decades, supported universities through dedicated academic programs. However, the ongoing expenses associated with maintaining EDA tool flows are not covered by research funding. Additionally, the costs for support staff necessary to operate the IT infrastructure are beyond the capabilities of many universities. 

Furthermore, MPW approaches have historically enabled universities to access advanced technologies. While this strategy has been valuable in the past, it is becoming increasingly difficult to sustain with the progression to very advanced technology nodes and the adoption of 3D integration strategies. The complexity and costs associated with these cutting-edge technologies make them less accessible through shared wafer programs. Accessing new technologies presents additional hurdles due to the stringent requirements imposed by foundries providing the necessary PDKs and libraries. For example, universities are required to have completed tape-outs in several previous node generations before being granted access to recent technologies. This prerequisite makes the process highly selective and eliminates the possibility of freely choosing foundries or exploring technologies. 
The requirements extend further to the necessity for fixed and fully detailed project descriptions accompanied by secured funding. Such stipulations effectively rule out free experimentation in research.
Multiple test-chip tape-outs are often required to determine the optimal technology selection for a given project. Typical research budgets do not allow for such extensive experimentation. Some PDKs and libraries necessitate installation in isolated environments separate from the university's IT infrastructure. Access is confined to secure locations on campus, which demands significant effort and resources to establish and maintain. 

In response to these limitations, some universities are turning to open-source PDKs, which offer a more accessible pathway to chip design education. Open-source PDKs can alleviate some barriers by providing free access without NDAs and stringent requirements. However, these solutions come with their own set of challenges. The availability of open-source PDKs depends on the goodwill of foundries, as they must publish the PDKs. Potential reliability issues may also arise, given that open-source PDKs may not be as rigorously validated as commercial ones. Furthermore, the current state of open-source PDKs supports only a limited set of older technologies, such as 180\,nm \cite{gf180-pdk} or 130\,nm nodes \cite{gf130-pdk}. While these nodes are sufficient for educational purposes, they are no suitable alternatives for chip design research that requires access to newer technology nodes. This restricts the scope of research and hinders innovations.

\subsection{Availability vs. Enablement} \label{subsec:availability-enablement}
While access to EDA tools, PDKs, libraries, and IP cores is crucial for chip design, their mere availability is insufficient. The significant challenges lie in the design enablement, encompassing multifaceted and resource-intensive tasks. They include the setup and maintenance of IT infrastructure, installation and updating of EDA tools, management of technology-specific databases such as PDKs, libraries, IP blocks, and generators (e.g., memory generators), and the technology-specific configuration of EDA tools. Moreover, automation of the design flow through scripting and the provision of user interfaces to ensure usability add further complexity.

For individual research groups or small academic institutions, managing the enablement tasks poses a significant challenge, often exceeding their technical and financial capacities. To alleviate this burden, cloud-based approaches are becoming increasingly popular within the semiconductor design community. These solutions offer scalable computing resources for chip design tasks and provide a single access point with centralized management of all enablement-related activities. Such platforms create a continuum between semiconductor technological developments and their applications within components or systems, including AI applications, sensors or network components. This need is further amplified by the advent of 3D communication substrates compatible with chiplets. The chiplet-based mix-and-match approach to system design requires interoperability and reusability, further increasing the overall design flow complexity. 

Examples of existing cloud platforms include open-source initiatives like TinyTapeout \cite{tinytapeout}, which deploys the OpenROAD design flow \cite{openroad} using GitHub Actions. Open-source EDA tools, IP and PDKs greatly simplify design enablement through unrestricted distribution and usability. However, open-source flows are not yet competitive with proprietary ones in terms of PPA metrics. On the commercial side, companies focusing on design enablement services offer platforms like makeChip by Racyics. Furthermore, to strengthen the European chip design ecosystem Europractice is leading efforts to establish a cloud-based design infrastructure \cite{europractice-cloud}. The development of cloud platforms for chip design is still in its early stages. The main drawback of existing solutions is that they currently cover only a limited set of technologies. Expanding the reach and capabilities of cloud-based enablement platforms remains a critical objective for the semiconductor design community to fully realize the benefits of collaborative and scalable design infrastructures. 

The challenges identified in Sections~\ref{subsec:recruiting} to \ref{subsec:availability-enablement} are also experienced by other countries such as the United States. The U.S. National Science Foundation (NSF) has sponsored several workshops aiming to understand the needs of academia in IC design and fabrication. The reports published based on these workshops indicate similar challenges~\cite{b4, nsf-eda-workshop}.

\section{Opportunities and Strategies for Improvement}

\begin{itemize}
    \item[\ding{228}] \textbf{Recommendation 1}: \textit{Low-barrier programs in schools.}
\end{itemize}
Pre-university initiatives and contests with the aim of fostering interest in chip design in schools have traditionally targeted top-performing students. While engaging these students is crucial, there exists a significant untapped potential among the broader student population. Students who may not initially excel academically could develop a strong interest and aptitude in chip design if given the opportunity and appropriate support.

To unlock this hidden potential, it is imperative to develop strategies that motivate a wider range of students, prevent discouragement, and ignite a passion for chip design. One effective approach is to lower the barriers to entry by introducing chip design concepts at higher levels of abstraction. This can be achieved through the use of HLS tools or leveraging Large Language Models (LLMs) for prompt-based hardware design. Integrating these methodologies with playful applications can further enhance student interest and participation.

Expanding and adapting existing educational platforms can also play a significant role. Platforms like TinyTapeout and FPGAWars \cite{fpgawars} offer valuable resources for school-age children but are often limited by language barriers, as most materials are in English. By translating these resources into the native languages of the children, educators can make chip design concepts more accessible and inclusive.

Online learning platforms and professional development courses for teachers may complement the above efforts. 

\begin{itemize}
    \item[\ding{228}] \textbf{Recommendation 2}: \textit{Information campaigns.}
\end{itemize}
Information campaigns in schools and universities are another imperative path motivating students to pursue specializations in chip design. One effective strategy is organizing industry visiting days and excursions. By facilitating visits to semiconductor companies and research labs, students gain firsthand experience of the chip design process and its real-world applications. Such experiences demystify the industry, making it more appealing to students who may be undecided about their specialization.

Establishing online information centers is another crucial recommendation. These platforms can provide comprehensive resources on potential career paths, presenting sub-fields of chip design such as RF, sensor, analog, or digital design, detailing what to study and offering insights into the day-to-day responsibilities of a chip designer. Presenting success stories of chip designers can capture the students' interest and imagination. Role models in chip design are especially important to target the persistent gender and diversity gap.

Creating online maps that display universities offering chip design programs can assist students in making informed decisions about their education. These maps can include details on curricula, research groups, and their specific focus areas, as well as highlighting nearby chip design industries. This geographical and program-specific information can help students choose institutions that best fit their career aspirations.

Finally, incorporating podcasts and video materials into the campaigns can greatly enhance their impact. Podcasts featuring interviews with industry experts or advice on career development can be both informative and inspiring. Educational videos that break down complex concepts into understandable segments can make chip design more accessible. These multimedia resources can be distributed through social media platforms.

\begin{itemize}
    \item[\ding{228}] \textbf{Recommendation 3}: \textit{Coordinated education funding.}
\end{itemize}
Innovative recruiting and outreach activities, such as those described above, can significantly benefit from the focused support of government funding agencies. Effectively coordinating efforts across multiple agencies within a country, across Europe, or on an international scale demands substantial organizational effort, strategic alignment, and collaboration. One such example effort is a recent solicitation, jointly announced by the U.S. NSF and Department of Commerce, for the establishment of a Network Coordination Hub for the National Network for Microelectronics Education (NNME)~\cite{nnme}. The Network Coordination Hub will oversee the establishment and operation of Regional Nodes, ensuring the delivery of consistent, high-quality educational offerings across the U.S. The Regional Nodes will provide rigorous and engaging curricula, instructional materials, experiential learning opportunities, professional development for educators, and more. This coordinated hierarchical structure aims to standardize and enhance educational experiences nationwide. 

\begin{itemize}
    \item[\ding{228}] \textbf{Recommendation 4}: \textit{Enhancement of automation and standardization.}
\end{itemize}
Both backend and frontend design productivity improve significantly with enhanced automation and standardization. To increase the efficiency of frontend design, it is necessary to raise the abstraction level of the hardware description. This can be achieved by improving existing tools and methods such as HLS or HCLs to reduce manual coding. Further, by utilizing new AI methods, such as LLMs, prompt-based tools can be developed that assist in generating and optimizing designs. 

Despite the disparate tools required in the backend chip design flow and its technology dependence, it is inherently structured into abstract steps. This natural partitioning presents an opportunity to streamline the design process. Vendor- and technology-independent templates representing the design steps could be a solution. Such a standardized framework can be adapted to various technologies and toolsets facilitating design enablement. Reference designs and flows also  contribute considerably to backend productivity and thus design enablement. 

\begin{itemize}
    \item[\ding{228}] \textbf{Recommendation 5}: \textit{Open-source hardware.}
\end{itemize}
Harnessing the collective expertise of the global community through open-source efforts can significantly amplify the impact of such tools and frameworks as described in Recommendation 4. Open-source development fosters collaboration, accelerates innovation, and promotes standardization across the industry. Open source EDA tools and PDKs eliminate the dependency on NDAs and vendor- or foundry-specific restrictions.

Further, the availability of open-source hardware IP is a key enabler. The main advantage of open-source IP is accessibility, and the freedom to instantiate and modify pre-designed building blocks for complex chips, thereby increasing productivity. However, high IP quality is extremely important, not only in terms of verification maturity, but also in terms of availability of collaterals (documentation, synthesis and simulation scripts, integration harness).

\begin{itemize}
    \item[\ding{228}] \textbf{Recommendation 6}: \textit{Strengthening of Europractice.}
\end{itemize}
Europractice is the cornerstone of academic chip design across Europe by providing universities with access to EDA tools, libraries, PDKs, and IP. Moreover, Europractice acts as a unifying platform that promotes cross-national cooperation, streamlining interactions with EDA companies and foundries and simplifying negotiations for access to new technologies.

Given its pivotal role, it is imperative that Europractice is not only maintained but further strengthened. The two most critical areas requiring attention are chip design enablement and the reduction of costs associated with MPW runs for universities. While the ongoing efforts to establish a cloud-based design platform within Europractice target the enablement, cost reduction for MPW runs is still an open issue. A possible solution could be a corporate sponsorship program akin to the Efabless Open MPW Program \cite{efabless} in the United States. Additionally, establishing industry funds wherein companies contribute to support academic chip design could provide a sustainable financial model. This approach not only eases the financial burden on research groups but also strengthens ties between academia and industry.

\begin{itemize}
    \item[\ding{228}] \textbf{Recommendation 7}: \textit{Establishment of centralized design enablement infrastructure.}
\end{itemize}
Establishing one or more centralized design enablement hubs can be a cost-effective solution to (i) lower the barriers to accessing EDA tools, PDKs, and IPs for students and academic researchers, and (ii) enable students at all appropriate levels to design IC chips, hence broaden participation in IC chip design beyond those institutions currently engaged in these activities. Such an infrastructure is best to be cloud-based and provides tiered access to the broad community. The infrastructure should support the entire IC chip design process  from behavioral/structural description at the RTL or above to GDSII fabrication mask file generation. Furthermore, it is critical for the infrastructure to  provide assistance or even one-stop-shop for licensing, access, and maintenance of both commercial and open-source EDA tools for design and verification, as well as PDK/IPs at various technology nodes. To support more advanced research, providing access to design and simulation tools for chips designed with emerging technologies is required. The U.S. NSF has announced such a solicitation in 2024~\cite{cdh}.
Europractice also plans to launch a cloud-based design enablement platform \cite{europractice-cloud}. 

Such infrastructure hubs can be enhanced by support from volunteering expert groups within academia. By leveraging the expertise of university staff and researchers, development efforts can be offloaded and accelerated. In this context, establishing an online forum for university staff generates substantial impact. Such a forum would facilitate the exchange of knowledge, resources, and best practices. Further, including volunteering experts from academia and template-based automation frameworks for the design flow --- as described in Recommendation 4 --- can accelerate the integration of new technologies.

\begin{itemize}
    \item[\ding{228}] \textbf{Recommendation 8}: \textit{Target group-oriented enablement strategies.}
\end{itemize}
In teaching chip design, a one-size-fits-all enablement solution is unlikely since the spectrum of learners ranges from high-school to PhD students. To foster inclusivity, it is imperative to implement enablement strategies for multiple experience levels. Advertising the possibilities and ensuring accessibility to resources at each stage of the learner's development is essential.

For beginners in chip design — such as high-school or early undergraduate students — the focus should be on simplicity and accessibility. Emphasizing easy-to-use design flows, low-cost fabrication options, and introductory learning resources is key. At this level, there is no need for advanced technology nodes or complex customization of design flows. An example of a recommended pathway is the TinyTapeout initiative.

At the intermediate level, including late bachelor’s or beginning master’s students, the educational strategy should shift towards a deeper understanding of design flow internals. Learners benefit from adapting and customizing design processes and gaining familiarity with industry-grade tools and services. Therefore, initiatives like the IHP OpenPDK \cite{ihp-pdk} combined with the OpenROAD design flow offer a comprehensive package.

For advanced learners — master’s thesis students and PhD candidates — the requirements become more specialized. Their work often necessitates the use of advanced technology nodes and demands access to commercial PDKs and EDA tools. In such cases, leveraging commercial enablement services is a possibility. In future, the Europractice cloud-based enablement infrastructure provides a viable option too.

\section{Conclusion}
In this paper, we analyzed the role of European universities and academic initiatives in enhancing chip design education and research to address deficits in Europe’s chip design capabilities. We identified critical challenges in recruiting, productivity, technology access, and design enablement, and proposed recommendations to overcome these challenges. Our analysis reveals that European universities possess significant untapped potential to not only address the skilled labor shortage in chip design but also pave the way for sustained technological innovation and economic growth. In the future, implementing the proposed strategies will require coordinated efforts between academia, governments, and industry stakeholders.

\section*{Acknowledgment}
We are grateful to Dr. X. Sharon Hu, a Program Director at the U.S. National Science Foundation, for sharing her perspectives on the NSF’s programs related to infrastructure and workforce development in chip design and design automation.

\end{document}